\documentclass[sigconf,screen]{acmart}

\usepackage{todonotes}
\theoremstyle{definition}
\newtheorem*{definition}{Definition}

\AtBeginDocument{%
  }

\begin{document}

\title{When 1+1 does not equal 2: Synergy in games}

\author{Joshua Kritz}
\email{j.s.kritz@qmul.ac.uk}
\orcid{0000000}
\affiliation{%
  \institution{Queen Mary University of London}
  \city{London}
  \country{United Kingdom}
}

\author{Raluca D. Gaina}
\email{r.d.gaina@qmul.ac.uk}
\affiliation{%
  \institution{Queen Mary University of London}
  \city{London}
  \country{United Kingdom}
}


\renewcommand{\shortauthors}{Kritz et al.}

\begin{abstract}

Although \textbf{synergy} is an important concept that is strongly ingrained in games, it has not been widely discussed by the games community. This is due to the vagueness of the concept and the fact that there is no clear agreement on what it means. To solve this, we present a strict definition of what is synergy. Then we propose a methodology to use this definition to analyze synergy in games. Applying this definition to various games (Chess, League of Legends, and Magic: The Gathering), we illustrate how it can be used to solve many of the practical problems related to synergy.
  
\end{abstract}




\maketitle

\section{Introduction}

\textit{Synergy} is a complex word. It is used in diverse contexts in which it has slightly different meanings. In fact, to the knowledge of the authors, there is not a single context in which synergy is used with a solid and widely agreed definition. Naturally, games are not free of this dilemma.

Games have always been related to synergy by the various parties and stakeholders involved. There are several mentions and discussions of what synergy is, how to implement it, where it happens in games, etc. Evidence of this can be found in many places, as further discussed in Section 2. They provide interesting insight on how gamers and game designers/developers view synergy. This discussion covers all types of games, from digital to analog, from strategy to action games. 






Synergy is an inseparable part of games, and anyone involved with games should give more attention to it. Gamers use it to enhance their experience, to break games, and try their best to exploit it. However, their view is only one possible angle on the subject. 

Scholars and designers should actively discuss synergy and its usage in games. Having a more firm grip on what synergy is and how it manifests in games allows for better game design, and its detailed study would take steps towards opening paths to studies of the topic. For example: game balance is a property of games studied in depth~\cite{schreiber2021game}; a better understanding of synergy enables analysis of how synergy feeds into balance and improves this area.

This work aims to help with this conundrum by providing a definition of synergy. We analyze existing discussions and condense the notion and understanding of what is synergy. We aim to allow for a more active, purposeful, and directed use of synergy in games.

We provide here the first steps in the research of what synergy is. Continuing this work will then allow us to better understand the topic and its challenges.








\section{Literature Review}

In a variety of informal discussions \cite{agoodusern4me_what_2022,notgayyy_how_2022,whymme_how_nodate}, where synergy is mentioned, the discussion degrades to what is synergy rather than staying within the original topic. Even when the subject is introduced with a clear meaning of synergy, it becomes a point of contention in the following discussion.

However, more experienced industry experts provide a hard definition of synergy. Rosewater~\cite{rosewater_living_2013} defines synergy as ``connecting things such that they produce something more potent in aggregate than in isolation''. On the other hand, the Game Maker Toolkit video~\cite{game_maker_toolkit_how_2024} uses synergy as ``when two or more elements combine to produce a more powerful effect than the sum of the individual elements''. Those definitions are good for illustrating the general consensus of what synergy is. They serve the purpose of their context and discussion, but are not extensible to all applications, and thus can be improved to be more flexible in their application.

The formal literature has different issues. Some authors, particularly shorter articles, do not define or provide any context for what they mean by synergy \cite{franca_creativestone_2024,lee_draftrec_2022}. This creates confusion when trying to understand motivation, goals, methodology, or even results.

Traditional books usually define synergy when they use it~\cite{salen_rules_2004}. However, to ensure that its meaning is clear, they go through lengthy explanations, comparisons, and correlations. This shows how complex this topic is, while also confirming how important it is to thoroughly discuss the meaning of synergy.

Nevertheless, most of these discussions of synergy are not only about what synergy is. Most of the time, synergy is mentioned within a context of other features or related to other concepts, rather than individually.

A common topic is how to achieve balanced synergy. Two reddit discussions explicitly ask this question \cite{garydallison_synergies_2023,notgayyy_how_2022}. The first counts on 20 replies with good value for the discussion (excluding short hate comments). The second counts with 15 responses with good contribution and citing 1 external source (a designer's blog post). In an RPG forum, three users discuss a similar question\cite{whymme_how_nodate}. Game designers also have videos on youtube on the topic of how balanced synergy is important to games~\cite{ruswick_importance_nodate,game_maker_toolkit_how_2024}. Magruder et al. \cite{magruder_conservative_2022} also talk about a part of balancing that is controlling power creep on Magic The Gathering (MTG). From all discourse, we can identify some insights:
\begin{itemize}
    \item Balancing synergy in practice is difficult and time-consuming.
    \item Synergy is an key factor for the quality of the game.
    \item Designing synergy is tricky, but essential to game design.
\end{itemize}

Mark Rosewater \cite{rosewater_living_2013} and Franca et al. \cite{franca_creativestone_2024} agree that synergy is important for player creativity. That is, it allows players to experience the game in their own way and express themselves through gameplay. This is relevant to the intrinsic motivation of players, which is an important factor players consider when choosing which game to play \cite{tyack_self-determination_2020}.
Within this, there are players who look specifically for games with intense levels of synergy \cite{noauthor_good_2021,noauthor_lack_nodate,viktrol13_games_2022}. As such, we can see that synergy is a defining factor in player experience.

A common interest between many discussions is to find a surefire method to create synergy or a system that ensures good synergy \cite{salen_rules_2004,whymme_how_nodate,notgayyy_how_2022,agoodusern4me_what_2022}. However, there is no agreement on the best way to \textit{create} synergies. Frequently, it is even stated that it is \textit{impossible} to ever have even guidelines on how to achieve these goals. We disagree with this statement, instead being one of the motives for promoting a better discussion on synergy, to eventually be able to provide such detailed guidelines.

\section{Synergy Definition}

To define what synergy is, we draw from the commonalities between existing conceptualizations.

First, it always involves two or more elements of a game. The word element is intentionally vague, as what is combined to create synergy can be of any type: rules, game pieces, actions, etc. 

Second, the elements involved need to have at least one relation of impact, that is, one element has an impact on another. This relation does not need to be reflexive. 

Third, a combination should always produce results different from the expected non-synergic combination. Most of the time, this is to explain results that are different from the sum of the elements present in the synergy. However, there are cases where using the word ``synergy", we want to define simply an alternative result to the one expected.

\begin{definition}[Synergy]
\textit{Synergy is a set made up of two or more game elements that have a measurable value and where the value of the set is different from the sum of the individual outcomes of its elements.}
\end{definition}

What \textit{value} means in this definition depends on the game analyzed. For example, a game where the objective is to acquire a higher score than the opponent, the value of such action is the score it leads to. However, some games, where actions have a cost, might define the value as the cost effectiveness (score divided by cost).

A different approach to value would be something along the lines of what Rosewater correlates: that synergy is creativity. If the objective is to measure creativity, there are a few innovation metrics that can be used to measure the \textit{creative value} of a set. Although calculating the creativity of a single individual is usually pointless, this synergy definition can be used to compare different sets and to see which set has more or less synergy than others. 
The need for this use case depends on the problem at hand.

The usage of this synergy definition should vary with the game and the objective of its analysis; however, we can point out some general steps:

\begin{enumerate}
    \item \textbf{Define the objective of the analysis}: For example, is it to find which sets have better synergy? Is it to identify outlier values? Or to evaluate whether there is any synergy at all within the game? 
    \item \textbf{Define elements that can synergize}: For example, in an Role Playing Game (RPG), when evaluating the synergy between skills and equipments, it would not make sense to include characters as possible elements.
    \item \textbf{Define the value of an element}. This could be associated score, victory points, damage, cost, etc. Then create a formula that can evaluate this for each element of the game and ensure that it can also calculate the value of a set of elements. This formula does not need to be continuous, or even numerical; it is possible to have a classification formula (e.g., freezing, cold, warm, hot).
    \item \textbf{Define the sets (combinations of elements) to be analyzed and calculate their \textit{synergy values}}. This will generate data that can be used to evaluate the original objective.
    \item \textbf{Review and iterate}. This is a point where the measurement formula or the value can be reviewed, whether it is because it did not answer all of the initial questions, or it feels unreliable. Finding the correct measurement of value is part of identifying what \textit{synergy} is in your game.
\end{enumerate}

Analyzing synergy is not a straightforward process, but rather an iterative investigation of the game features. It is important to use a mixed-initiative approach to apply this definition in practice, since it is not possible to define a single formula or algorithm that works for all cases, and human designer input is critical. 

\section{Case Studies}

\begin{figure}
    \centering
    \includegraphics[trim={0 3.5cm 0 3.5cm},clip,width=0.9\linewidth]{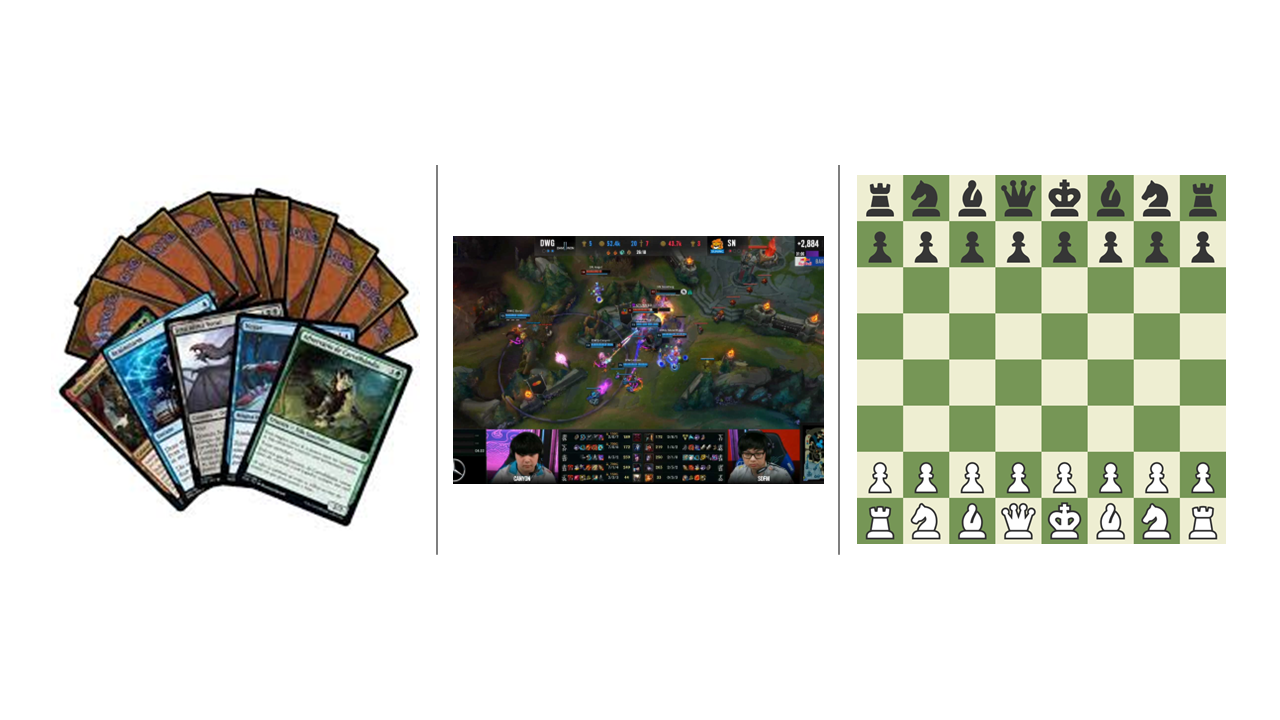}
    \caption{Case studies games - Magic: The Gathering (left), League of Legends (middle), Chess (right).}
    \label{fig:usecase}
\end{figure}

To better illustrate how we can use this method to analyze games, we will show its application in several existing games (see Figure~\ref{fig:usecase}).

\subsection{Chess}
We start with a very traditional game: Chess. Intuitively, chess is known to have no synergy. Nevertheless, it is possible to try to find some level of synergy within the game. A challenge with Chess is that you only ever use one piece within a turn, hence it is not intuitive how to identify what would be a set of game elements that could yield synergy.

Here, our objective would be to evaluate whether there is any synergy in the game. Since we cannot use multiple pieces in a turn, we can attempt to use a sequence of piece usage as a synergy set, e.g. using the knight after the bishop, or the knight after the rook.

In terms of value, most chess metrics are only concerned with the outcome of the game, so we could use changes in the win rate. Another common proxy for the game state is evaluating the total value of the remaining \textit{pawn} pieces of a player (a knight is worth three pawns, a queen nine). The second option does not fit our definition of the synergy set, as the sequence of usage of pieces does not have a direct impact on this value, and hence we cannot measure it. Using the first option, we can evaluate all possible sequences of piece usages and calculate their win rates. 

However, chess is a perfect information game that has been studied by professionals for years. Those studies led to what is called \textit{openings} which are a set of predetermined moves from the start of the game. They define a play pattern that defines the progress of the game as they are made under the assumption that the opponent is making the optimal plays. These openings are considered strategies rather than synergies, so we could use the same logic for the synergy sets we defined. Here we could iterate and try to find another way to define synergy sets to avoid this issue. Or we can accept that there is no such synergy set, and then, by our definition, Chess indeed has no synergy.


The challenge presented by Chess to this methodology is that it is difficult to define what elements can synergize in the game. We have to know which elements can interact in a possible synergy to define what the value is and how to measure it.

\subsection{Character-Based Games}
On another spectrum, there are genres of games in which it is widely agreed that there is a lot of synergy. Two examples are Multiplayer Online Battle Arenas (MOBA), like League of Legends and Dota 2, and Hero Shooters, like Valorant or Overwatch. 

A common feature of all of them is that the players choose different characters to play in each match. These characters have different powers and characteristics that make them unique from each other. There is a great deal of interest in the synergy between these characters because it is an important feature for the prediction of a match win \cite{hodge_win_2021} and character recommendation \cite{lee_draftrec_2022}, both extensive areas of research. 

The synergy between the characters can be seen from two angles: a) positive synergy, when two characters work well \textit{with} each other, or b) negative synergy, when they do not go well \textit{against} each other (frequently named \textit{counters} or \textit{counter-picks}). The main difference between them is that the first looks at the relationship between characters on the same team, whereas the second correlates characters in opposite teams. 

An example of positive synergy in the game League of Legends is as follows: Anivia is a character that has the ability to create a wall, and Vayne has the skill to push opponents and stun them if they hit a wall. Thus, Anivia can create more circumstances for Vayne to make the most of her push ability. As for negative synergy, or counter: Kindred creates a ground buff that protects allies inside this area, and Gragas has a skill that pushes all enemies away from a target point. This makes it easy for Gragas to push Kindred's allies outside of the buff area.

Let us focus on the first case. 
There are many outcomes that can be analyzed, but, most of the time, we are interested in finding out if the characters have positive synergy, that is, if they have higher win rates when in the same team. Furthermore, our application for this analysis will be to recommend characters for a player given the current characters selected in their team.

We can use several metrics to evaluate this. The most straightforward approach is to look at historic game data and calculate their win rate when on the same team. 
Another more player-centric approach would be to use player opinions/experience on whether they are good together if these data are available. 
A third idea would be to evaluate the results of team fights where both are present versus when they are not; 

All three metrics can have their use in a recommendation system, but if we want to maximize player success, the first metric is more reliable. An interesting characteristic of these metrics is that they will capture the negative synergy between characters, too, that is, when they directly hinder each other and naturally include it in the recommendation system. This highlights the need for a mixed-initiative approach as the choice of the best metric needs human input.

The possible sets we want to analyze are restricted, in the sense that we don't need to calculate all possible character combinations. We have a limited set of existing teammates to evaluate, so we will need to calculate the synergy between all possible characters and the currently selected ones.

Some improvements can be made in this example. We can analyze the synergy between more than two characters within the same team or include the possible choices of players who did not choose their own characters. This will increase the number of calculations that must be performed. From another angle, we can look at the counter-picks, calculating the synergy between possible characters and the characters present in the enemy team.


\subsection{Trading Card Games}

Trading Card Games (TCG) and Collectible Card Games (CCG) are genres of games where players collect cards and use them to first build a deck, which then they use to play the game. These games have a large pool of different cards: Magic The Gathering (MTG), the oldest of them, has over 20 thousand different playable cards. Such games have a high degree of synergy, they live off the synergy between the cards that exist within the game \cite{rosewater_living_2013}, and this is the reason why players play these games.

TCGs are built with synergy in mind, and the cards are created to interact with each other. In MTG, we have cards like \textit{Master of the pearl trident} that increase the power of other creatures of the type \textit{merfolk}; this is clearly designed to be played with and synergize well with other merfolk cards. In Hearthstone, \textit{Warsong Commander} buffs minions you play with power 3 or less, and this card is made to be played with such minions. There are also synergies that are less straightforward and sometimes not intentional. In the first example, the master of the pearl trident also gives the \textit{islandwalk} keyword to other merfolks; that is, they benefit from the opponent having an island, but, if your opponent has no islands, it does nothing. A card named \textit{Spreading seas} transforms an opponent's card into an island, and it is very strong with the islandwalk keyword. 


The main point of TCGs is that they have a life cycle in which the game releases new sets of cards periodically to increase the variety of available cards and, by extension, new synergies and strategies. These sets often contain hundreds of new unique cards. Testing and balancing all these new cards against the existing ones is the biggest challenge for designing TCGs, as it is necessary to find extremely strong synergies that dominate others and create a stale gameplay.


The value we are interested in evaluating is the \textit{strength} of a card. To calculate this, we could try an evaluation analogous to the MOBA example, looking at the win rates of the cards and combinations of cards. However, a new set of cards does not have gameplay data, making it impossible to calculate win rates. Thus, we need to use a formula that only considers the intrinsic values of the game. Since the objective of the game is to reduce the opponent life total to 0, let us use the amount of damage a card gives as the value for its strength. An important part of the game is also its resource system, mana, which is quite relevant to the impact of a card on the game, so it is better to use the ratio of \textit{damage per mana spent} as a strength measurement.

The possible combinations we need to test are more than just calculating the synergy between every pair of cards. Since decks in MTG can contain up to 60 cards with up to 4 copies of each, and throughout the game players play multiple cards, we need to account for combinations of many cards and even consider combinations with the same card featured more than once. Given that every new set brings over 100 new cards and thousands of existing cards, this will create a number of possible synergy sets on the order of $10^{115}$, which is higher than the number of atoms in the universe. 

If we calculate these millions of combinations, we will have a very accurate result of overly powerful synergies. Designers can then fine-tune these cards to bring them more in line with the other synergies. If they need to make many changes, it would be best to recalculate everything with the new versions of the cards until we are assured that all cards are balanced.

Although calculating all possible combinations will definitely solve the problem, this is rarely possible. The number of possible combinations of cards in a single deck increases exponentially with each released card. This is the biggest challenge for TCGs regardless of the methodology followed for balance. 

\section{Conclusion}

This work presents the importance of synergy, a concept that is closely related to games in all forms. All stakeholders interested in games should pay more attention to synergy. However, because of the elusive nature of the concept, most avoid discussing it or have a hard time doing so.

To solve this, we present a formal and strict definition of synergy. We then exemplify how this definition fits in various games and how flexible it is in its application.

Empowered by this definition, stakeholders in games can discuss synergy more concretely and can even use it to analyze games. This is particularly important for game design and development, where balancing synergy has been a challenge for a long time.

\section{Future Work}

The next step is to turn to an important stakeholder, the game designers. We want to identify precisely how they deal with synergy in their game design process and the points where they have a hard time dealing with it. Then we need to present this definition and ask how it can be incorporated into the design process.

Here, we present a theoretical application of this synergy definition, using a tentative methodology brought about by intuition. Therefore, an important next step is defining a technical implementation of this process and applying it to various games to investigate any shortcomings when widely used in practice. This aims to build trust with stakeholders in this definition and then promote its wide use for better game design.


\begin{acks}
 This work was supported by the EPSRC Centre for Doctoral Training in Intelligent Games \& Games Intelligence (IGGI) EP/S022325/1.
\end{acks}

\bibliographystyle{ACM-Reference-Format}
\bibliography{references,other}

\end{document}